\documentclass[draftclsnofoot, 12pt, onecolumn]{IEEEtran}

\usepackage{graphicx}
\usepackage{color}
\usepackage{placeins}
\usepackage{float}
\usepackage{tabularx,colortbl}
\usepackage{amsmath}
\usepackage{multirow}
\usepackage{hyperref}

\begin{document}
%
\title{On Phase Noise Suppression in Full-Duplex Systems}

\author{Elsayed Ahmed, and Ahmed M. Eltawil\footnote{Elsayed Ahmed and Ahmed M. Eltawil are with the Department of Electrical Engineering and Computer Science at the University of California, Irvine, CA, USA (e-mail: \{ahmede, aeltawil\}@uci.edu).}
}

\maketitle

\vspace{-0.5in}

\begin{abstract}
Oscillator phase noise has been shown to be one of the main performance limiting factors in full-duplex systems. In this paper, we consider the problem of self-interference cancellation with phase noise suppression in full-duplex systems. The feasibility of performing phase noise suppression in full-duplex systems in terms of both complexity and achieved gain is analytically and experimentally investigated. First, the effect of phase noise on full-duplex systems and the possibility of performing phase noise suppression are studied. Two different phase noise suppression techniques with a detailed complexity analysis are then proposed. For each suppression technique, both free-running and phase locked loop based oscillators are considered. Due to the fact that full-duplex system performance highly depends on hardware impairments, experimental analysis is essential for reliable results. In this paper, the performance of the proposed techniques is experimentally investigated in a typical indoor environment. The experimental results are shown to confirm the results obtained from numerical simulations on two different experimental research platforms. At the end, the tradeoff between the required complexity and the gain achieved using phase noise suppression is discussed.
\end{abstract}

%
\IEEEpeerreviewmaketitle

\section{Introduction}
Due to the tremendous increase in wireless data traffic, full-duplex transmission was introduced as a promising duplexing mechanism that could potentially double the spectral efficiency of wireless systems. The main limitation impacting full-duplex transmission is managing the strong self-interference signal imposed by the transmit antenna on the receive antenna within the same transceiver. To double the spectral efficiency, full-duplex systems should be able to mitigate the self-interference signal to below the receiver noise floor. Throughout the literature, several combinations of self-interference cancellation schemes have been proposed ~\cite{Ref1}-\cite{Ref12} aiming to mitigate the self-interference signal below the receiver noise floor. However, several experimental results~\cite{Ref1}-\cite{Ref7} have demonstrated that, using conventional hardware, complete self-interference elimination is highly challenging, mainly due to a combination of hardware imperfections, especially radio circuits' impairments.

In order to identify the system limitations, several recent publications~\cite{Ref6}-\cite{Ref8} have considered the problem of self-interference cancellation in full-duplex systems to investigate the impact of radio circuit impairments on the cancellation capability. More specifically, in~\cite{Ref8} an analytical model that includes transmitter and receiver phase noise, quantization noise, and receiver AWGN noise is introduced. The results show that among the mentioned three impairments, transmitter and receiver oscillator phase noise is the main bottleneck that limits self-interference mitigation capability. In~\cite{Ref6} an experimental framework using the Wireless Open-Access Research Platform (WARP)~\cite{Ref14} is used to investigate the main cause of performance bottlenecks in current full-duplex systems. The results show that the key bottleneck in current systems is the phase noise of the transmitter and receiver local oscillators. In addition to phase noise, other experimental results~\cite{Ref7} show that when using very low phase noise transceivers (e.g. test instruments), transmitter nonlinearity becomes the main performance limiting factor. As a conclusion, among the various radio frequency (RF) circuits' impairments, oscillator phase noise and transmitter nonlinearities are found to be the main self-interference cancellation limiting factors in full-duplex systems.

Giving it limiting impact on performance, phase noise reduction is one of the main design targets for full-duplex system designers. Phase noise could be reduced by either designing high quality, low phase noise oscillators (which is highly impractical ) or by performing phase noise estimation and suppression. In this work, we analytically and experimentally investigate the problem of phase noise estimation and suppression in full-duplex systems in the presence of both transmitter and receiver oscillator phase noise. First, we study the impact of oscillator phase noise on full-duplex orthogonal frequency division multiplexing (OFDM) systems. For practical system considerations, both free-running and phase locked loop (PLL) based oscillators are considered. Second, in addition to the frequency-domain technique proposed by the authors in~\cite{Ref13}, we propose another reduced complexity time-domain phase noise estimation and suppression technique. Detailed complexity comparison between the two proposed techniques is introduced. Third, the effect of channel estimation error on the phase noise estimation performance is discussed. Fourth, a real-time experimental framework is used to confirm the conclusions derived from the numerical analysis. For additional diversity, the experimental results are obtained using two different research platforms (e.g. WARP~\cite{Ref14}, and USRP~\cite{Ref15}). Finally, the overall system performance is investigated to study the feasibility of using phase noise estimation and suppression techniques in full-duplex systems in terms of achieved gain and required complexity.

Generally, the presence of phase noise in OFDM systems introduces common phase error (CPE) and intercarrier interference (ICI)~\cite{Ref16}-\cite{Ref17}. Most of the current self-interference cancellation schemes compensate only for the CPE and ignore the ICI effect, which limits the amount of cancellable self-interference power to the ICI power level. Therefore, improving self-interference cancellation capability requires the ICI signal to be estimated and suppressed. In fact, conventional half-duplex ICI suppression techniques~\cite{Ref16}-\cite{Ref18} could be used in full-duplex systems with the following two exceptions; first, in full-duplex systems, while suppressing the ICI associated with the self-interference signal, the signal-of-interest is considered as an unknown noise signal. Second, in full-duplex systems, the self-interference signal is known at the receiver side, thus eliminating the need to use decision feedback techniques to obtain the transmitted signal.

Since the ICI suppression amount depends on the accuracy of the estimated ICI signal, which is proportionally related to the computational complexity; the main challenge in full-duplex systems is achieving sufficient ICI suppression at reasonable computational complexity. The analysis shows that in full-duplex systems, two main factors affect the achieved ICI suppression amount; first, the fact that the signal-of-interest is considered as a noise signal during the ICI estimation process significantly degrades the quality of the estimated ICI signal, especially in the cases where the signal-of-interest power is higher than the ICI signal power. Second, the results also show that using different oscillator types (e.g. free-running or PLL based oscillator) affect the achieved ICI suppression amount, mainly due to the different phase noise power spectral density shapes in different oscillator types.

In full-duplex systems, typically analytical and numerical analysis are not sufficient for accurate conclusions; mainly due to the significant dependence of the performance on hardware impairments, which are very challenging to accurately model. Thus, numerical analysis confirmed by experimental analysis is the best way to study full-duplex systems. In this work, following the numerical analysis, an experimental framework with different research platforms is used to investigate the performance of the proposed phase noise estimation and suppression techniques in typical indoor environments. The experimental results are shown to confirm the conclusions derived from the numerical analysis.

The remainder of this paper is organized as follows. In Section II, the signal model is presented. The proposed phase noise estimation and suppression techniques are introduced in Section III. Analysis and discussions are presented in Section IV. Finally, section V presents the conclusion.

\emph{Notation}: We use $(*)$ to denote convolution, $(.)^H$ to denote conjugate transpose, $E[.]$ to denote expectation. We use boldface letters $(\textbf A)$ for matrices, $\textbf A(m,n)$ to denote the element on the $m^{th}$ row and $n^{th}$ column of the matrix $\textbf A$, and $diag(\textbf A)$ to denote a diagonal matrix whose diagonal is constructed from the vector $\textbf A$.

\section{Signal Model}
In this section, a signal model for full-duplex systems including the transmitter and receiver phase noise is introduced. For sufficient self-interference suppression in full-duplex systems, one or more self-interference cancellation schemes should be used. Self-interference cancellation schemes could be divided into three main categories: (i) passive suppression, (ii) RF cancellation, and (iii) digital cancellation. In passive suppression, the self-interference signal is suppressed in the propagation domain before it is processed by the receiver circuitry. Passive suppression could be achieved using antenna separation and/or shielding~\cite{Ref1}-\cite{Ref2}, directional antennas~\cite{Ref9}-\cite{Ref10}, or careful antenna placement~\cite{Ref3}. In RF cancellation~\cite{Ref7}, a copy of the transmitted RF signal is used to mitigate the self-interference signal at the low-noise amplifier input. In digital cancellation techniques~\cite{Ref1,Ref7,Ref8}, the self-interference signal is canceled in the digital domain by leveraging the fact that the transceiver knows the signal it is transmitting. Typically, digital cancellation is used to remove the residual self-interference after passive or RF cancellation.
\begin{figure}[ht]
\begin{center}
\noindent
  \includegraphics[width=4in ,height=2.0in]{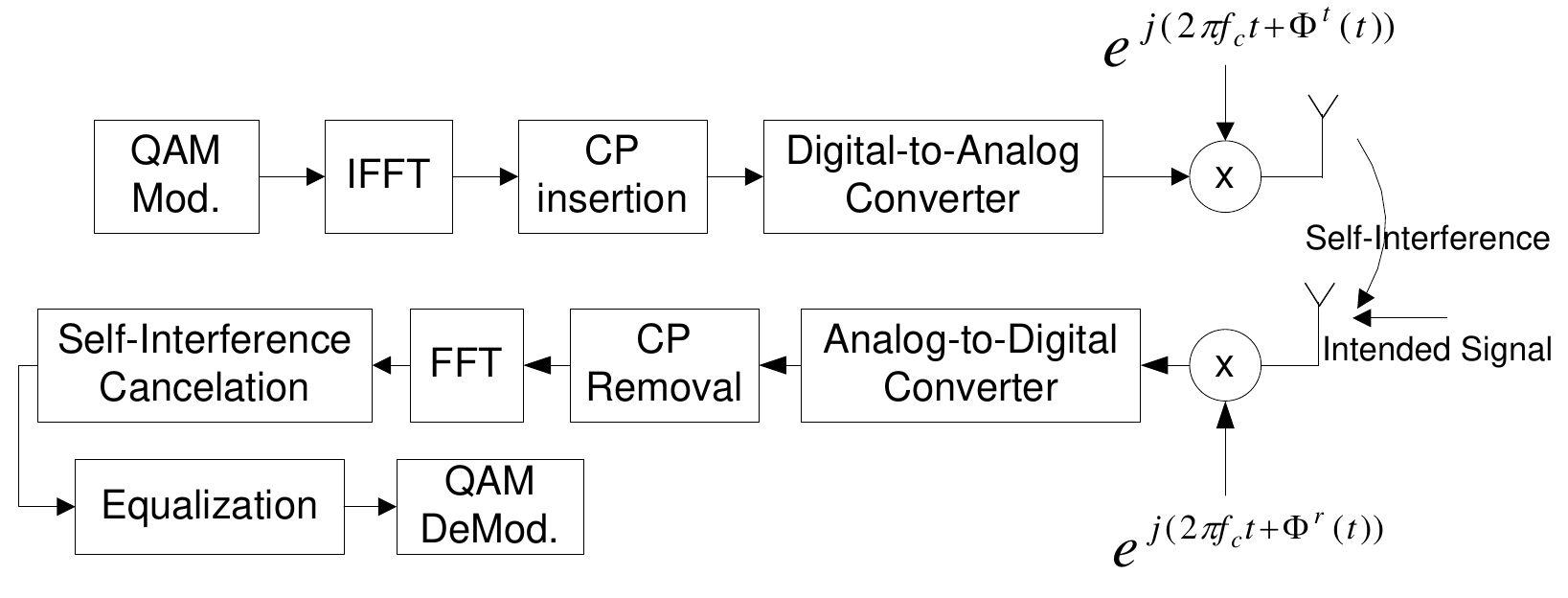}
  \caption{Block diagram of full-duplex OFDM transceiver.\label{Fig1Label}}
\end{center}
\end{figure}

Figure~\ref{Fig1Label} illustrates a block diagram for a full-duplex OFDM transceiver using passive suppression followed by digital self-interference cancelation. At the transmitter side, the base-band signal is modulated using an OFDM modulator and then up-converted to the carrier frequency $f_c$. The oscillator at the transmitter side is assumed to have a random phase error represented by $\phi^t(t)$. At the receiver side, the received signal consists of the self-interference (the signal from the transmitter in the same transceiver) and the signal-of-interest (the signal to be decoded) down-converted from the carrier frequency to the base-band. The down-conversion mixer is assumed to have a random phase error represented by $\phi^r(t)$. The base-band signal is then converted to the frequency domain using Fourier transform. In the frequency domain, the self-interference signal is estimated and subtracted from the received signal. Finally, the output of the self-interference cancellation block is equalized and demodulated to restore the transmitted data.

The received base-band time domain signal can be written as
\begin{equation}\label{eq:1}
y_n = \left[\left(x_n^I e^{j\phi_n^{t,I}} * h_n^I \right) + \left(x_n^S e^{j \phi_n^{t,S}} * h_n^S \right)\right] e^{j \phi_n^r} + z_n\text{,}
\end{equation}
where $n$ is the sample index, $x^I$, $x^S$ are the transmitted self-interference signal and signal-of-interest respectively, $\phi^{t,I}$, $\phi^{t,S}$ are the self-interference and signal-of-interest transmitter phase noise processes, $\phi^r$ is the receiver phase noise process, $h^I$, $h^S$ are the self-interference and signal-of-interest channels, and $z$ is the receiver noise. 

Performing Discrete Fourier Transform (DFT) on both sides of~\eqref{eq:1} we get
\begin{eqnarray}\label{eq:2}
Y_k &=& \underbrace{\sum_{m=0}^{N-1} \sum_{l=0}^{N-1} X_l^I H_m^I J_{m-l}^{t,I} J_{k-m}^r}_{Y_k^I} + \underbrace{\sum_{m=0}^{N-1} \sum_{l=0}^{N-1} X_l^S H_m^S J_{m-l}^{t,S} J_{k-m}^r}_{Y_k^S} + Z_k \nonumber \\
&=& Y_k^I+Y_k^S+Z_k\text{,}
\end{eqnarray}
where $k$ is the subcarrier index, $N$ is the total number of subcarriers per OFDM symbol, $Y_k^I$, $Y_k^S$ represents the self-interference and signal-of-interest parts of the received signal, $Z_k$ is the Fourier transform of the receiver noise, and $J^i$, $i\in{[(t,I),(t,S),r]}$ represents the DFT coefficients of the phase noise signal calculated as
\begin{equation}\label{eq:3}
J_k^i = \sum_{n=0}^{N-1} e^{j\phi_n^i} e^{-j2\pi nk/N}\text{.}
\end{equation}
In experimental results published in~\cite{Ref3r,Ref4} it was shown that for full-duplex systems with a strong self-interference line-of-sight component, the self-interference channel follows a Rician distribution with a very large Rician factor (e.g. 25dB to 35dB), and thus can be considered as a frequency-flat channel over wide frequency bands. In the analysis portion of this paper, for simplicity, we assume a frequency-flat channel while developing the signal model. Accordingly, Equation~\eqref{eq:2} can be simplified as
\begin{equation}\label{eq:4}
Y_k = \sum_{l=0}^{N-1}X_l^I H_l^I \sum_{m=0}^{N-1}J_{m-l}^{t,I} J_{k-m}^r + Y_k^S+Z_k = \sum_{l=0}^{N-1}X_l^I H_l^I J_{k-l}^c + Y_k^S+ Z_k\text{,}         
\end{equation}
where $J^c$ is the DFT coefficients of the combined transmitter and receiver phase noise calculated as the circular convolution of $J^{t,I}$ and $J^r$.

Rewriting~\eqref{eq:4} in a more detailed form we get
\begin{equation}\label{eq:5}
Y_k = X_k^I H_k^I \underbrace{J_0^c}_{CPE} + \underbrace{\sum_{l=0,l\neq k}^{N-1}X_l^I H_l^I J_{k-l}^c}_{ICI} + Y_k^S+Z_k\text{,}
\end{equation}
where $J_0^c$ is the DC coefficient that acts on all subcarriers as a CPE, and the second term represents the ICI associated with the self-interference signal. The time-domain representation of~\eqref{eq:5} can be written as
\begin{equation}\label{eq:5r}
y_n = (x_n^I*h_n^I) j_n^c + y_n^S+z_n\text{,}
\end{equation}
where $j_n^c = e^{j\left(\phi_n^{t,I}+\phi_n^{r}\right)}$ is the time domain representation of the combined phase noise process.

\begin{figure}[ht]
\begin{center}
\noindent
  \includegraphics[width=3in ,height=1.3in]{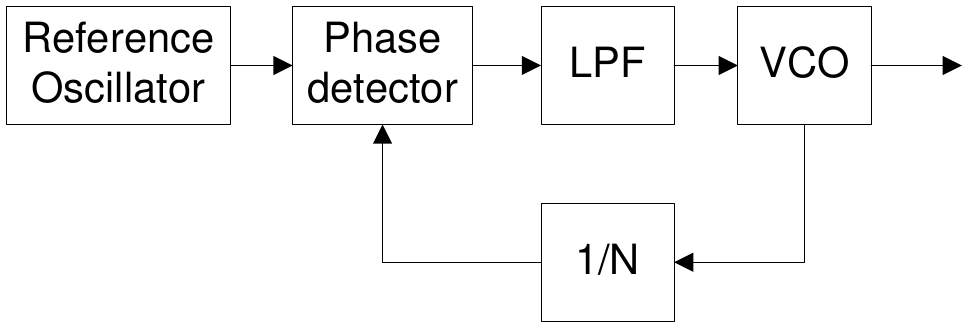}
  \caption{PLL based oscillator.\label{Fig2Label}}
\end{center}
\end{figure}

In order to proceed with the analysis, a closed form model for the phase noise process is required.  In this paper, we consider the two commonly used oscillator types; free-running oscillators and PLL based oscillators. 
%
%
In free-running oscillators the phase noise could be modeled as a Wiener process~\cite{Ref19} where the phase error at the $n^{th}$ sample is related to the previous one as $\phi_n = \phi_{n-1} + \alpha$, where $\alpha$ is a Gaussian random variable with zero mean and variance $\sigma^2=4\pi^2f_c^2CT_s$. In this notation $T_s$ describes the sample interval and $C$ is an oscillator dependent parameter that determines its quality. The oscillator parameter $C$ is related to the 3dB bandwidth $f_{3dB}$ of the phase noise Lorentzian spectrum by $C=f_{3dB}/\pi f_c^2$. As shown in~\cite{Ref16}, the phase noise auto-correlation for free-running oscillators is calculated as
\begin{equation}\label{eq:6}
E\left[e^{j\phi_m}e^{-j\phi_n}\right] = E\left[e^{j\Delta\phi_{mn}}\right] = e^{\frac{-4\pi^2 f_c^2 C T_s |m-n|}{2}}\text{.}
\end{equation}
%
%
In PLL based oscillators, as shown in figure~\ref{Fig2Label}, the voltage controlled oscillator (VCO) output is controlled through a feed-back loop that involves a phase detector and low-pass filter (LPF). The purpose of the feed-back loop is to lock the phase of the VCO output with the phase of a high quality reference oscillator. As shown in~\cite{Ref20}, the PLL output phase noise can be modeled as Ornstein-Uhlenbeck process with auto-correlation function calculated as
\begin{equation}\label{eq:7}
E\left[e^{j\Delta\phi_{mn}}\right] = e^{\frac{-4\pi^2 f_c^2}{2}\left(C T_s |m-n| + 2\sum_{i=0}^{n_0}\left(\mu_i+v_i\right)\left(1-e^{-\lambda_i T_s |m-n|}\right)\right)}\text{,}
\end{equation}
where ($n_0$, $\mu$ , $v$, $\lambda$) are PLL specific parameters that are function of the PLL loop filter design\footnote{See~\cite{Ref20} for detailed description on how these parameters are calculated.}.

\section{Self-Interference cancellation with phase noise suppression}
According to~\eqref{eq:5}, total self-interference cancellation requires both the CPE and the ICI components to be suppressed. Conventional digital self-interference cancellation schemes only consider the suppression of the CPE component and neglects the ICI component, which limit the amount of cancellable self-interference power to the ICI power level. In this section, we introduce two different phase noise estimation and suppression techniques that are used to enhance self-interference cancellation capability in full-duplex systems.

Generally, self-interference cancellation requires the knowledge of both transmitted self-interference signal ($X^I$) and self-interference channel ($H^I$). Since it is transmitted from the same transceiver, the transmitted self-interference signal is assumed to be known at the receiver side. An accurate estimation for the self-interference channel ($H^I$) as well as the signal-of-interest channel ($H^S$) could be obtained using orthogonal training sequences sent at the beginning of each transmission frame.
\subsection{Frequency-domain phase noise estimation and suppression}
In this technique, the DFT coefficients of the phase noise process ($J_k$) is estimated in the frequency-domain, and then used to suppress both the CPE and ICI components. The estimation-suppression process consists of four main steps;
\begin{itemize}
\item
Estimating the DC coefficient ($J_0^c$).
\item
Suppressing the CPE component by subtracting $X_k^I H_k^I J_0^c$ from the received signal.
\item
Estimating the remaining phase noise coefficients ($J_i^c, i\neq 0$).
\item
Suppressing the ICI component by reconstructing the signal $\sum_{l=0,l\neq k}^{N-1}X_l^I H_l^I J_{k-l}^c$ then subtract it from the received signal.
\end{itemize}

For the DC coefficient estimation, the least square (LS) estimator is used as follows
\begin{equation}\label{eq:8}
J_0^c = \frac{1}{N_u} \sum_{k=0,k\in U}^{N_u-1}\frac{Y_k}{X_k^I H_k^I} \text{,}         
\end{equation}
where $U$ is a set that contains the pilot positions within the OFDM symbol, and $N_u$ is the number of pilot subcarriers. After estimating the DC coefficient, the CPE component is subtracted from the received signal in~\eqref{eq:5} as follows
\begin{equation}\label{eq:9}
Y_k-X_k^I H_k^I J_0^c=\sum_{l=0,l\neq k}^{N-1}X_l^I H_l^I J_{k-l}^c +Y_k^S+Z_k\text{.}         
\end{equation}

In order to perform ICI suppression, the remaining coefficients of $J^c$ have to be estimated. Based on~\eqref{eq:9}, the problem of estimating $J^c$ is considered as a linear estimation problem, where $J^c$ is a parameter vector distributed by Gaussian noise and the signal-of-interest ($Y^S$). For an estimation order $M$ (where $M$ is the number of coefficients to be estimated), Equation~\eqref{eq:9} can be written in a matrix form as
\begin{equation}\label{eq:10}
\left[ \begin{array}{c} B_{l_1} \\ B_{l_2} \\:\\ B_{l_p} \end{array} \right] =
\begin{bmatrix} A_{l_1} & ... &A_{l_1+M} \\ A_{l_2} & ... &A_{l_2+M}\\:&:&: \\ A_{l_p} & ... &A_{l_p+M} \end{bmatrix}
\left[ \begin{array}{c} J_{M/2}^c \\:\\J_1^c\\J_-1^c\\:\\ J_{-M/2}^c \end{array} \right] 
+ \left[ \begin{array}{c} Y^S_{l_1} \\ Y^S_{l_2} \\:\\ Y^S_{l_p} \end{array} \right] 
+ \left[ \begin{array}{c} \gamma^{ICI}_{l_1} \\ \gamma^{ICI}_{l_2} \\:\\ \gamma^{ICI}_{l_p} \end{array} \right] 
+ \left[ \begin{array}{c} Z_{l_1} \\ Z_{l_2} \\:\\ Z_{l_p} \end{array} \right]\text{,} 
\end{equation}
where $B_k=Y_k-X_k^I H_k^I J_0^c$ , $A_k=X_k^I  H_k^I$, and $\gamma^{ICI}$ is the residual ICI beyond the estimation order $M$. The set $[l_1 \ \ l_2 \ \ ... \ \ l_p]$ has to be of length $\geq M$ in order to solve~\eqref{eq:10} for $M$ unknowns. Summarizing~\eqref{eq:10} in a compact form we get
\begin{equation}\label{eq:11}
\bf B=\bf A \bf J^c + \boldsymbol{\eta} \text{,}         
\end{equation}
where $\boldsymbol{\eta}$ represents the effective noise that combines all of the signal-of-interest, the residual ICI, and the receiver noise. Using~\eqref{eq:11}, the minimum mean square error (MMSE) estimate of $\bf J^c$ is given by
\begin{equation}\label{eq:12}
\bf J^c=\bf W \bf B\text{,}         
\end{equation}
\begin{equation}\label{eq:13}
\bf W=\bf R_{JJ} \bf A^H (\bf A \bf R_{JJ} \bf A^H + \bf R_{\eta \eta})^{-1} \text{,}         
\end{equation}
where $\bf R_{JJ}$ represents the correlation matrix of the vector $\bf J^c$, and $\bf R_{\eta \eta}$ represents the correlation matrix of the vector $\boldsymbol{\eta}$.

Using equation~\eqref{eq:3}, the (p,q) element of the correlation matrix $\bf R_{JJ}$ can be calculated as
\begin{equation}\label{eq:14}
\bf{R_{JJ}} (p,q)= E\left[J_p J_q^*\right] = \frac{1}{N^2}\sum_{m=0}^{N-1}\sum_{n=0}^{N-1}E\left[e^{j\Delta\phi_{mn}}\right]e^{-j\frac{2\pi}{N}(pm-qn)}\text{,} 
\end{equation}
where $E\left[e^{j\Delta\phi_{mn}}\right]$ is calculated as in~\eqref{eq:6},~\eqref{eq:7} for free-running and PLL based oscillators.

Assuming that the data symbols and the receiver noise are not correlated, the correlation matrix $\bf R_{\eta \eta}$ can be written as 
\begin{equation}\label{eq:15}
\bf R_{\eta \eta} = \mathsf{diag}(E\left[\left|Y_{l_1}^S \right|^2\right]+E\left[\left|\gamma_{l_1}^{ICI}\right|^2\right]+\sigma_z^2 ,........, E\left[\left|Y_{l_p}^S \right|^2\right]+E\left[\left|\gamma_{l_p}^{ICI}\right|^2\right]+\sigma_z^2)\text{,}         
\end{equation}
where $\sigma_z^2$ is the receiver noise variance, $E\left[\left|\gamma_{l_i}^{ICI}\right|^2\right]$ is the power of the residual ICI at subcarrier $l_i$ calculated as [13]
\begin{equation}\label{eq:16}
E\left[\left|\gamma_{l_i}^{ICI}\right|^2\right]=\sum_{p=0,p>|M|}^{N-1} \bf{R_{JJ}} (p,p)\text{,}         
\end{equation}
and $E\left[\left|Y_{l_1}^S\right|^2\right]$ is the power of the received signal-of-interest at subcarrier $l_i$. For simplicity, $E\left[\left|Y_{l_1}^S\right|^2\right]$ can be approximated to the average received signal-of-interest power as follows
\begin{equation}\label{eq:17}
E\left[\left|Y_{l_i}^S\right|^2\right]=E\left[\left|X_{l_i}^S H_{l_i}^S\right|^2\right] = E\left[\left|H_{l_i}^S\right|^2\right]\text{,}         
\end{equation}
where the transmitted signal are assumed to be M-QAM modulated with a unity average power. At the end, the phase noise vector $\bf J^c$ is constructed by placing the $M$ estimated coefficients in their corresponding positions and placing zero elsewhere.

In the ICI cancellation phase, the ICI component is reconstructed as
\begin{equation}\label{eq:18}
ICI_k = \sum_{l=0,l\neq k}^{N-1} X_l^I H_l^I J_{k-l}^c \text{,}         
\end{equation}
and then subtracted from the received signal.

Regarding the computational complexity, the most computation consuming part in the discussed frequency-domain technique is the calculation of the weighting matrix $\bf W$ which involves matrix inversion. Although the correlation matrix $\bf R_{JJ}$ is a symmetric matrix, however, due to the fact that $\bf A$ is a general matrix with no special properties, the matrix $(\bf A \bf R_{JJ} \bf A^H + \bf R_{\eta \eta})$ is also a general $M$x$M$ matrix, which sets the complexity of such technique to $O(M^3)$. The high complexity order will limit the use of such technique to small $M$ cases which directly affects the accuracy of the estimated phase noise vector and thus the amount of suppressed ICI power.

In addition to matrix inversion, reconstructing the ICI component involves a convolution process of order $O(NM)$, which also limits the use of such technique to systems with a small number of subcarriers $N$. For complexity reduction, a lower complexity time-domain phase noise estimation and suppression technique is proposed in the following subsection.

\subsection{Time-domain phase noise estimation and suppression}
refering back to~\eqref{eq:5r}, since the self-interference channel $H^I$ and the self-interference signal $X^I$ are known, a time-domain MMSE estimator could be used to solve~\eqref{eq:5r} for the unknown vector $j^c$ that is distributed by Gaussian noise and the signal-of-interest ($y^S$). For an $M$ order MMSE estimator, equation~\eqref{eq:5r} can be written in a matrix form as
\begin{equation}\label{eq:19}
\left[ \begin{array}{c} y_{l_1} \\ y_{l_2} \\:\\ y_{l_p} \end{array} \right] =
\begin{bmatrix} a_{l_1} & 0 & ... & 0 \\ 0 & a_{l_2} & ... & 0\\:&:&: \\ 0 & 0 & ... a_{l_p} \end{bmatrix}
\left[ \begin{array}{c} j_{M/2}^c \\:\\j_1^c\\j_-1^c\\:\\ j_{-M/2}^c \end{array} \right] 
+ \left[ \begin{array}{c} y^S_{l_1} \\ y^S_{l_2} \\:\\ y^S_{l_p} \end{array} \right] 
+ \left[ \begin{array}{c} z_{l_1} \\ z_{l_2} \\:\\ z_{l_p} \end{array} \right]\text{,} 
\end{equation}
or in a compact form as
\begin{equation}\label{eq:20}
\bf y=\bf a \bf j^c + \boldsymbol{\zeta} \text{,}         
\end{equation}
where $a_n=x_n^I*h_n^I$, $\boldsymbol{\zeta}$ represents the effective noise vector that combines Gaussian noise $\bf z$, and signal-of-interest $\bf y^S$. The set $[l_1 \ \ l_2 \ \ ... \ \ l_p]$ has to be of length $\geq M$ in order to solve~\eqref{eq:20} for $M$ unknowns.

Comparing~\eqref{eq:20} with~\eqref{eq:11} we note that in the time-domain technique, the matrix $\bf a$ is a diagonal matrix and the noise vector $\boldsymbol{\zeta}$ contains only the Gaussian noise $\bf z$, and the signal-of-interest $\bf y^S$. However, in the frequency-domain technique, $\bf A$ is a full matrix and the noise vector $\boldsymbol{\eta}$ has an additional term ($\gamma^{ICI}$) which is the residual ICI beyond the estimation order $M$. The reason is that the phase noise is a multiplicative process in the time-domain which means that the received signal at time $n$ is only affected by the phase noise at that time instant, while, in the frequency-domain, due to the ICI effect, the received signal at each subcarrier is affected by the phase noise at that subcarrier and all other subcarriers.

Using~\eqref{eq:20}, the MMSE estimate of $\bf j^c$ is given by
\begin{equation}\label{eq:21}
\bf j^c=\bf w \bf y\text{,}         
\end{equation}
\begin{equation}\label{eq:22}
\bf w=\bf R_{jj} \bf a^H (\bf a \bf R_{jj} \bf a^H + \bf R_{\zeta \zeta})^{-1} \text{,}         
\end{equation}
where $\bf R_{jj}$ represents the time-domain correlation matrix of the vector $\bf j^c$, and $\bf R_{\zeta \zeta}$ represents the correlation matrix of the vector $\boldsymbol{\zeta}$. The (m,n) element of the correlation matrix $\bf R_{jj}$ can be calculated as
\begin{equation}\label{eq:23}
\bf{R_{jj}} (m,n)= E\left[j_m j_n^*\right] = E\left[e^{j\Delta\phi_{mn}}\right]\text{,}
\end{equation}
where $E\left[e^{j\Delta\phi_{mn}}\right]$ is calculated as in~\eqref{eq:6},~\eqref{eq:7} for free-running and PLL based oscillators respectively. The noise correlation matrix $\bf R_{\zeta \zeta}$ can be calculated as in~\eqref{eq:15} with the exception that $\boldsymbol{\gamma}^{ICI} = \bf 0$ in the time-domain problem. 

In cases where $M<N$ (the number of estimated samples is less than the overall number of samples per OFDM symbol), the remaining un-estimated samples could be set to the value of one. However, for better estimation quality, the estimated samples are linearly interpolated to get an estimate for the phase noise at the un-estimated time positions. For lower interpolation errors, the estimation positions $[l_1 \ \ l_2 \ \ ... \ \ l_p]$ are chosen to be equally-spaced in the time-domain. For the cancellation phase, the self-interference signal is reconstructed in the time-domain as $y_n^I = \left(x_n^I*h_n^I \right)j_n^c$, then subtracted from the received signal.

From a performance perspective, two main advantages make the time-domain technique expected to outperform the frequency-domain technique; first, in the frequency-domain technique, the remaining ICI beyond the cancellation order is considered as a noise term that negatively affects the estimation quality. However, in the time-domain technique there is no ICI effect. Second, the linear interpolation performed in the time-domain technique results in a good estimate for the un-estimated samples beyond the estimation order, thus improving the overall estimation performance.

In terms of complexity, the complexity of the proposed time-domain technique is at least one order of magnitude lower than the complexity of the frequency-domain technique. In more details, the complexity advantage of the time-domain technique is a result of three main factors; first, In the calculation of the weighting matrix ($\bf w$), the time-domain phase noise correlation matrix $\bf R_{jj}$ is a real symmetric matrix and a is a diagonal matrix which results in $(\bf a \bf R_{jj} \bf a^H + \bf R_{\zeta \zeta})^{-1}$ being a symmetric matrix with an inversion complexity order $O(M^2)$ instead of $O(M^3)$ in the frequency-domain technique. Second, in the cancellation phase of the time-domain technique, the self-interference signal is reconstructed using $O(N)$ multiplication process instead of the $O(NM)$ convolution process used in the frequency-domain technique. Finally, performing time-domain phase noise interpolation helps to achieve better performance at a lower estimation order ($M$), and thus reduces the complexity. 

On the other hand, performing time-domain MMSE estimation requires additional inverse discrete fourier transform (IDFT) process to calculate $a_n=IDFT(X_k^I H_k^I)$. For $N=2^m$, the complexity of the IDFT process is $O(N\log_2N)$. Performing time-domain phase noise interpolation is an additional $O(N)$ process that does not exist in the frequency-domain technique. As a conclusion, table~\ref{Table1Label} summarizes the computational complexity of both time-domain and frequency-domain techniques.
\begin{table}[ht]
\caption{Complexity order of time-domain and frequency-domain phase noise estimation and suppression techniques}
\label{Table1Label}
\centering
\begin{tabular}{|c|c|c|}
\hline
  & Frequency-domain technique  & Time-domain technique \\
\hline
Estimation Phase & $O(M^3)$ & $O(M^2)$ For matrix inversion + \\
 & & $O(N\log_2N)$ to calculate $a_n=IDFT(X_k^I H_k^I)$ \\
\hline
Cancellation Phase & $O(NM)$ & $O(N)$ for Interpolation + \\
 & & $O(N)$ for Cancellation \\
\hline
\end{tabular}
\end{table}

\section{Analysis and discussions}
In this section, the performance of the proposed phase noise estimation and suppression techniques is experimentally and numerically investigated under different operating conditions. Both free-running and PLL based oscillators are considered. In addition, the effect of channel estimation error on performance is also investigated. The performance of the proposed techniques is compared to the case where no phase noise suppression is performed, showing the potential gain achieved by using such phase noise suppression techniques. Following the analysis, the feasibility of using phase noise estimation and suppression techniques in full-duplex systems is discussed in terms of achieved gain and required complexity.

A 20MHz wireless LAN system is used as a framework for the analysis. The system is assumed to operate in a full-duplex mode, where the wireless terminals are transmitting and receiving at the same time, using the same carrier frequency. The transmitted frame consists of orthogonal training sequences used for channel estimation purposes, followed by data OFDM symbols with 64 subcarriers in each symbol. Each OFDM symbol contains 4 pilot subcarriers used for CPE estimation. The carrier frequency $f_c$ is set to 2.4GHz with a system bandwidth of 20MHz. The indoor TGn channel model D~\cite{Ref21} is used to model the self-interference and signal-of-interest channels. The self-interference and signal-of-interest channel's Rician factors are set to 30dB and 3dB respectively.

At the receiver side, the orthogonal training sequences transmitted at the beginning of each frame are used to obtain an estimate for the self-interference as well as the signal-of-interest channels. The channel is estimated at each training symbol, and then averaged over all training symbols to obtain a more accurate channel estimate. The channel is estimated once per frame, and assumed to be constant within the frame duration.

In the analysis, the self-interference cancellation gain is used as a performance metric. Self-interference cancellation gain is defined as the incoming self-interference power divided by the remaining self-interference power after performing all cancellation and suppression processes. Generally, self-interference cancellation gain is used as a measure of how much self-interference suppression does the cancellation technique achieve. For more clarity, following is the definition of terms used in this section; 1) self-interference to signal-of-interest ratio (ISR) is defined as the ratio between the incoming self-interference power and the signal-of-interest power. 2) Total phase noise induced ICI power ($P_{ICI}$) is defined as the total power of the ICI component relative to the received self-interference power in dBc units. As an example, an ISR of $-$40dB and $P_{ICI}$ of $-$30dBc means that the signal-of-interest and the ICI component are below the self-interference signal by 40dB and 30dB respectively, it also means that the ICI power is higher than the signal-of-interest power by 10dB.

\subsection{Time-domain Vs frequency-domain phase noise suppression}
In this subsection, the proposed reduced complexity time-domain phase noise estimation and suppression technique is compared to the frequency-domain technique. A free-running oscillator and exact channel knowledge are assumed in this analysis. However, the PLL based oscillator and channel estimation error effects are studied separately in the following subsections.
\begin{figure}[!ht]
\begin{center}
\noindent
  \includegraphics[width=4in ,height=3in]{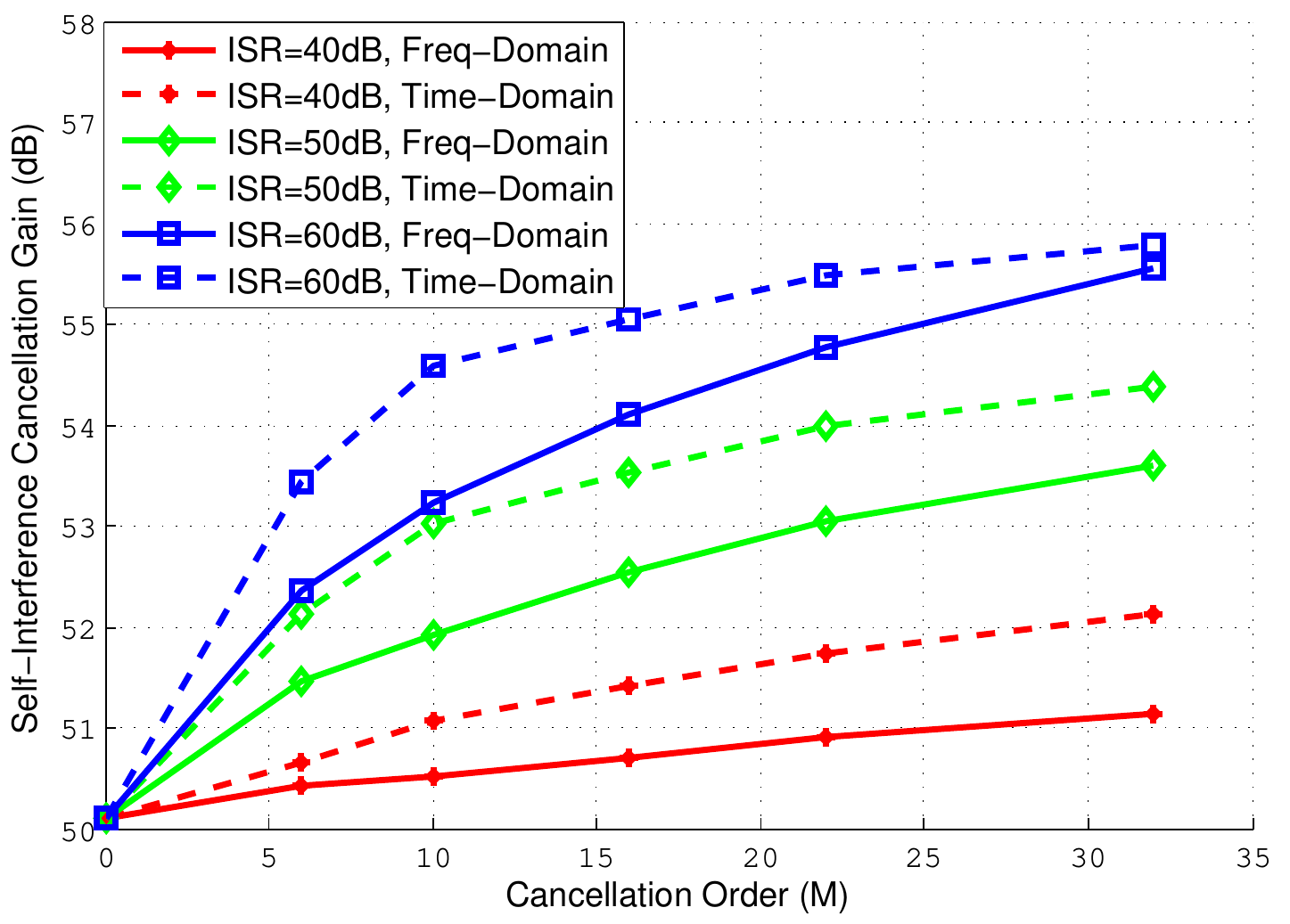}
  \caption{Self-interference cancellation gain for time- and frequency-domain phase noise estimation and suppression techniques at $P_{ICI} = -50$dBc, with free-running oscillator.\label{Fig3Label}}
\end{center}
\end{figure}

Figure~\ref{Fig3Label} shows the performance of time- and frequency-domain techniques at different cancellation orders ($M$), and different ISR values with a total ICI power of $-$50dBc. It has to be noticed that $M=0$ means that only the CPE component is suppressed, and no ICI suppression is performed. In fact, suppressing the CPE component only is exactly what most of the existing digital self-interference cancellation schemes are doing. The conclusions from this analysis are multifold; first, as described in section III, performing only CPE suppression limits the amount of cancellable self-interference power to the ICI power level (i.e. $-$50dBc in our case). Second, according to~\eqref{eq:11},~\eqref{eq:20} the variance of the noise vectors $\boldsymbol{\eta}$ and $\boldsymbol{\zeta}$ are directly proportional to the signal-of-interest power, therefore, increasing the signal-of-interest power (i.e. decreasing the ISR) increases the estimator noise variance, thus degrading the estimator performance. Finally, in addition to its complexity advantage, the proposed time-domain technique achieves better performance ($\sim$1dB more cancellation gain) compared to the frequency-domain technique. The performance superiority of the time-domain technique is mainly due to the linear interpolation performed using the estimated samples to get an estimate for the remaining samples in each OFDM symbol.
\subsection{Free-running Vs PLL based oscillators}
As a matter of fact, most of the current wireless system transceivers use PLL based oscillators; mainly due to its phase stability compared to the continuous phase drift in free-running oscillators. In order to understand the effect of using PLL based oscillators on the proposed phase noise estimation and suppression techniques, first we investigate the main differences between free-running and PLL based oscillators.

\begin{figure}[!ht]
\begin{center}
\noindent
  \includegraphics[width=4in ,height=3in]{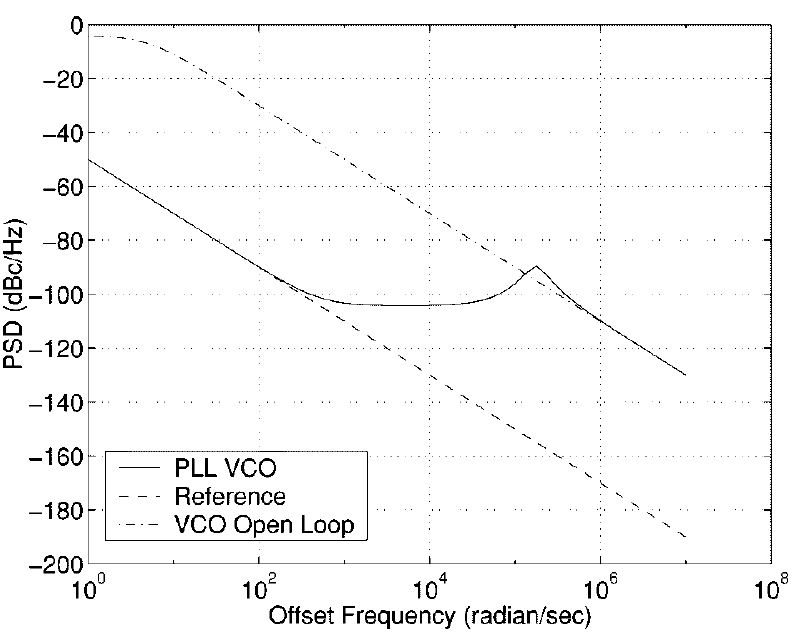}
  \caption{PSD for free-running and PLL based oscillators~\cite{Ref20}, free-running is the dashed lines and PLL is the solid line\label{Fig4Label}}
\end{center}
\end{figure}

\begin{figure}[!ht]
\begin{center}
\noindent
  \includegraphics[width=4in ,height=3in]{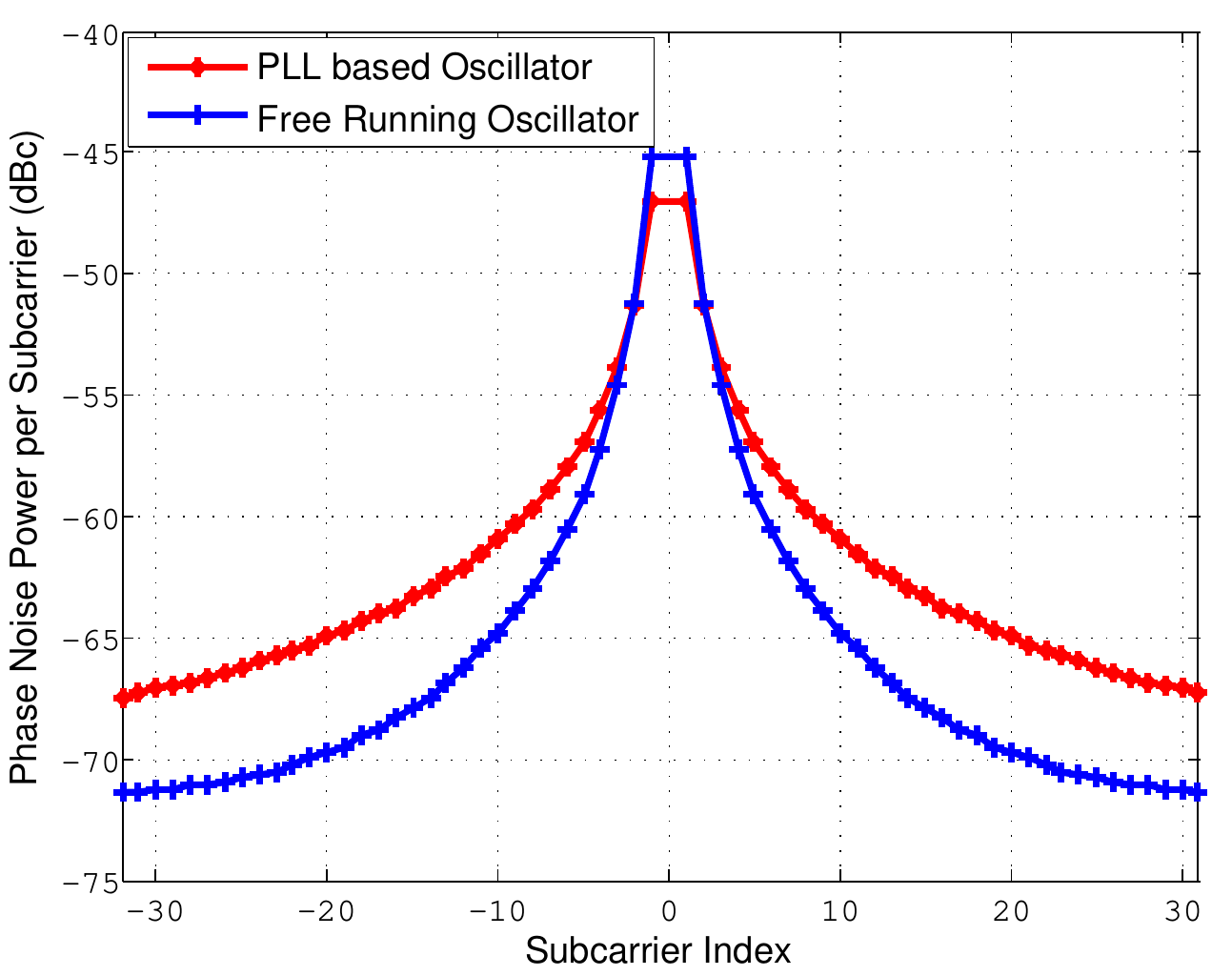}
  \caption{Phase noise power per subcarrier for free-running and PLL based oscillators at $P_{ICI}=-$40dBc.\label{Fig5Label}}
\end{center}
\end{figure}

First, in free-running oscillators the phase error is modeled as a Wiener process~\cite{Ref19} with a continuous phase drift. However, in PLL based oscillators~\cite{Ref20}, the feed-back loop tends to stabilize the output phase error which results in very small CPE compared to free-running oscillators. Therefore, in case of using PLL based oscillators the CPE estimation could be omitted or estimated over long time periods. Second, as shown in~\cite{Ref19,Ref20}, generally the phase noise power spectral density (PSD) has a low pass shape with a decay rate proportional to $1/f_o^2$ ($f_o$ is the frequency offset from the main carrier). However, in PLL based oscillators, due to the existence of the loop filter, the phase noise PSD of the output Flatter over the bands of interest. Figure~\ref{Fig4Label}~\cite{Ref20} shows a typical example for the phase noise PSD in free-running and PLL based oscillators\footnote{In Figure~\ref{Fig4Label}~\cite{Ref20}, the legend "Reference" and "VCO Open Loop" refers to free-running oscillators and "PLL VCO" refer to PLL based oscillator.}. From an OFDM perspective, in the free-running oscillator case, the subcarriers centered around the carrier frequency will have large phase noise power, and this power will decay fast when you go towards the edge subcarriers. On the other hand, in PLL based oscillator case, the phase noise power starts at a lower value and decays slower than free-running oscillators. Figure~\ref{Fig5Label} shows the phase noise power per subcarrier for both free-running and PLL based oscillators with the same total in-band phase noise power. The question to consider is how does this impact the performance of the proposed techniques?

\begin{figure}[ht]
\begin{center}
\noindent
  \includegraphics[width=6.5in ,height=3in]{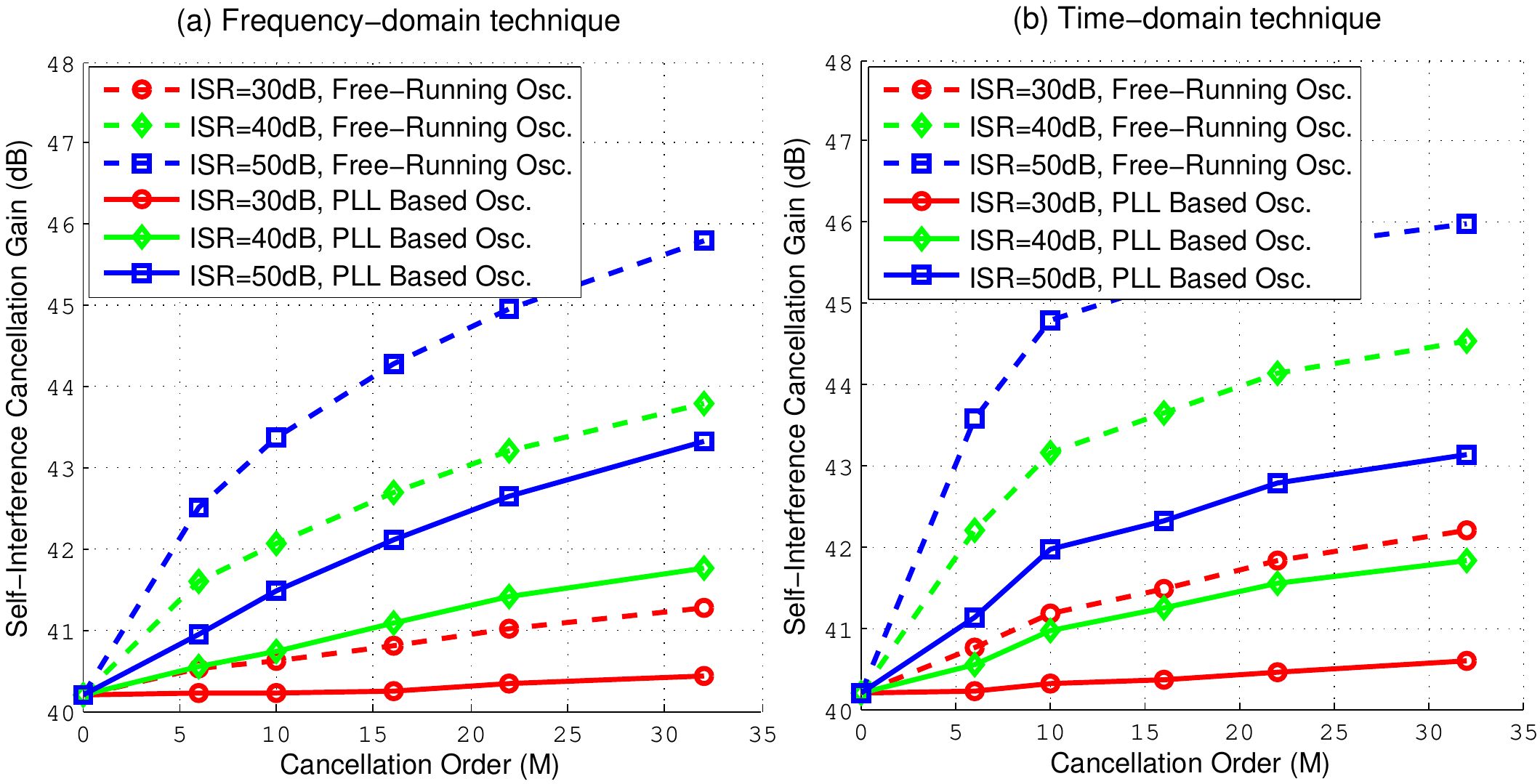}
  \caption{Self-interference cancellation gain for free-running and PLL based oscillators using frequency- and Time-domain phase noise estimation techniques at $P_{ICI} = -40$dBc.\label{Fig6Label}}
\end{center}
\end{figure}

In the frequency-domain estimation technique, at a given estimation order ($M$ arround the DC subcarrier), the power of the estimated phase noise will be larger in the case of free-running than PLL based oscillator. Therefore, the ICI suppression amount will be higher in the case of free-running oscillators. In addition, for small cancellation orders ($M$), the remaining ICI power beyond the estimation order will be smaller in case of free-running as compared to the PLL based oscillator, which means lower noise variance, and thus better estimation quality in the case of free-running oscillators. As a conclusion, using PLL based oscillators degrades the overall cancellation performance. Figure~\ref{Fig6Label}a shows the performance of the frequency-domain estimation technique using free-running and PLL based oscillators. The results show that using free-running oscillators approximately doubles the achieved ICI suppression amount compared to the case where PLL based oscillators are used.
%
%
%

For the time-domain estimation technique, the flatness in the phase noise PSD is equivalent to lower correlation between time-domain samples. Since the time-domain estimation technique uses linear interpolation to get an estimate for the remaining phase noise samples beyond the estimation order, lower time correlation means higher interpolation errors and thus lower estimation quality. Figure~\ref{Fig6Label}b shows the performance of the time-domain estimation technique using free-running and PLL based oscillators. The results also show that using PLL oscillators degrades the ICI suppression performance. On the other hand, comparing figure~\ref{Fig6Label}a and~\ref{Fig6Label}b we notice that even with PLL based oscillators, the time-domain technique still outperforms the frequency-domain technique.

\subsection{Effect of channel estimation error}
\begin{figure}[!ht]
\begin{center}
\noindent
  \includegraphics[width=4in ,height=3in]{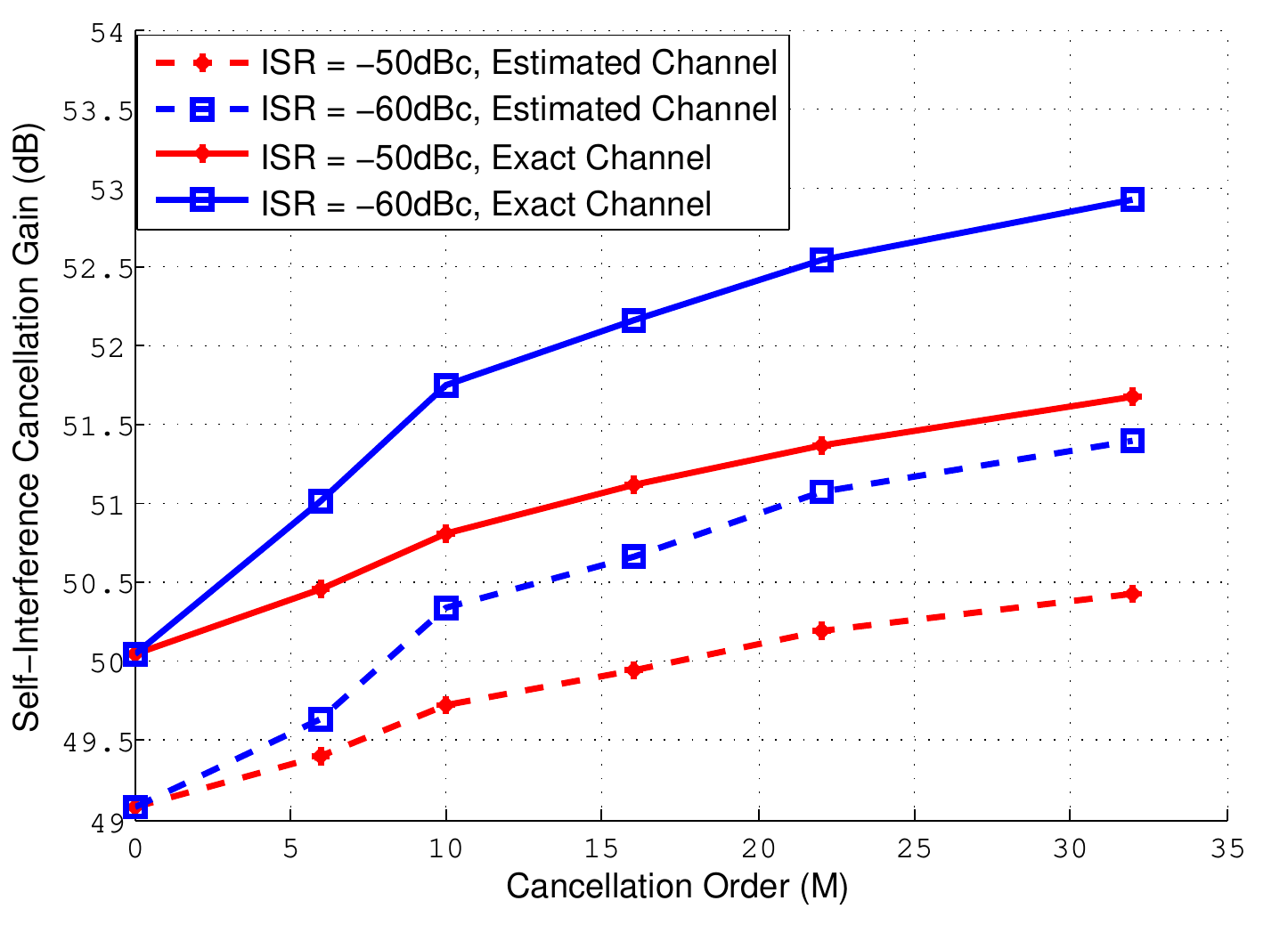}
  \caption{Self-interference cancellation gain for PLL based oscillators with time-domain phase noise estimation technique at $P_{ICI} = -50$dBc, using estimated and exact channels.\label{Fig8Label}}
\end{center}
\end{figure}

In practical systems, exact channel information is not available at the receiver side. Rather, the channel has to be estimated, which typically results in some channel estimation errors added to the received signal. In this analysis, we numerically investigate the effect of the channel estimation on the self-interference cancellation performance. Since the self-interference channel is involved in the estimation of both CPE and ICI components, the overall cancellation performance will be negatively affected by the channel estimation error. Figure~\ref{Fig8Label} shows the effect of the channel estimation error on the overall cancellation performance. The results show that channel estimation errors could results in a total of $\sim$1.5dB loss in the overall cancellation performance. It has to be noticed that, first, the performance of the ICI suppression technique is degraded by only 0.5dB and the other 1dB arises due to degradation in the estimation of the CPE component (compare the performance loss at $M=0$ and $M=32$ in figure~\ref{Fig8Label}). Second, at least 1dB performance degradation exists even with no ICI suppression (case with $M=0$), which means that conventional self-interference cancellation techniques are also affected by the channel estimation errors. Finally, despite the channel estimation errors, performing ICI suppression still improves the overall cancellation performance.
\subsection{Experimental analysis}
In full-duplex systems numerical analysis is necessary to investigate all different aspects and alternatives. However, numerical analysis is not sufficient mainly due to the lack of accurate modeling of hardware impairments and wireless channels, especially in the near-field where transmit and receive antennas within the same transceiver are placed. For that reason, performing experimental analysis is essential in full-duplex systems.

\begin{figure}[!ht]
\begin{center}
\noindent
  \includegraphics[width=5in ,height=2in]{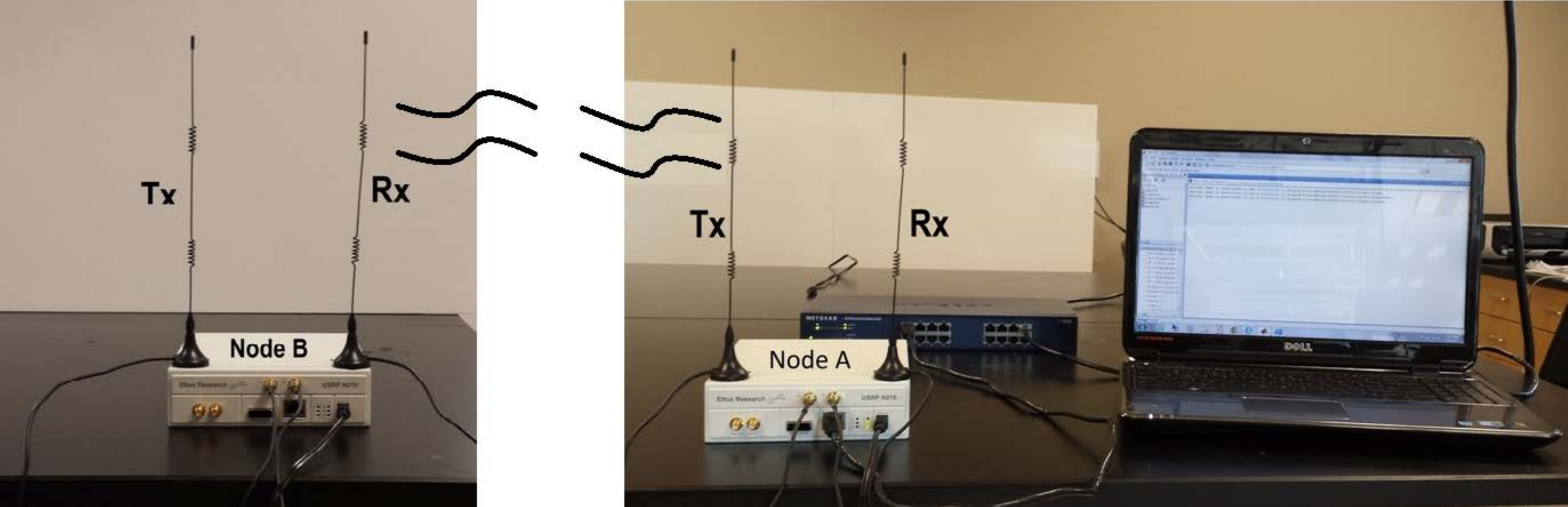}
  \caption{Experimental setup using USRP platform.\label{Fig9Label}}
\end{center}
\end{figure}

In this paper, a real time experimental frame work was constructed to investigate the performance of the proposed phase noise estimation and suppression techniques. For analysis diversity, two commonly used open access wireless platforms (e.g. WARP and USRP) are used in the analysis. Figure~\ref{Fig9Label} shows the experimental setup, where a full-duplex communication link is established using two research platforms namely node-A and node-B. Each node has one transmitter and one receiver connected to a separate antenna\footnote{note that one antenna and circulator could be used instead of two antennas.}. The transmitter base-band data is created using a PC and transferred to the platform through Ethernet cable. The platform converts the base-band data to pass-band and sends it over the air. At the receiver side, the platform receives the data and down-converts it to base-band. The down-converted data is then processed by the PC to obtain the results. Both platforms are configured to transmit and receive at the same time using the same carrier frequency. Based on the datasheet, both WARP and USRP platforms are using PLL based oscillator for the up- and down-conversion process. In the experimental analysis, the performance is evaluated at different ISR ratios. At the beginning of each transmission frame, orthogonal training sequences are sent for channel estimation proposes. The training sequences are also used to measure the ISR ratio at the receiver input.

\begin{figure}[!ht]
\begin{center}
\noindent
  \includegraphics[width=4in ,height=3in]{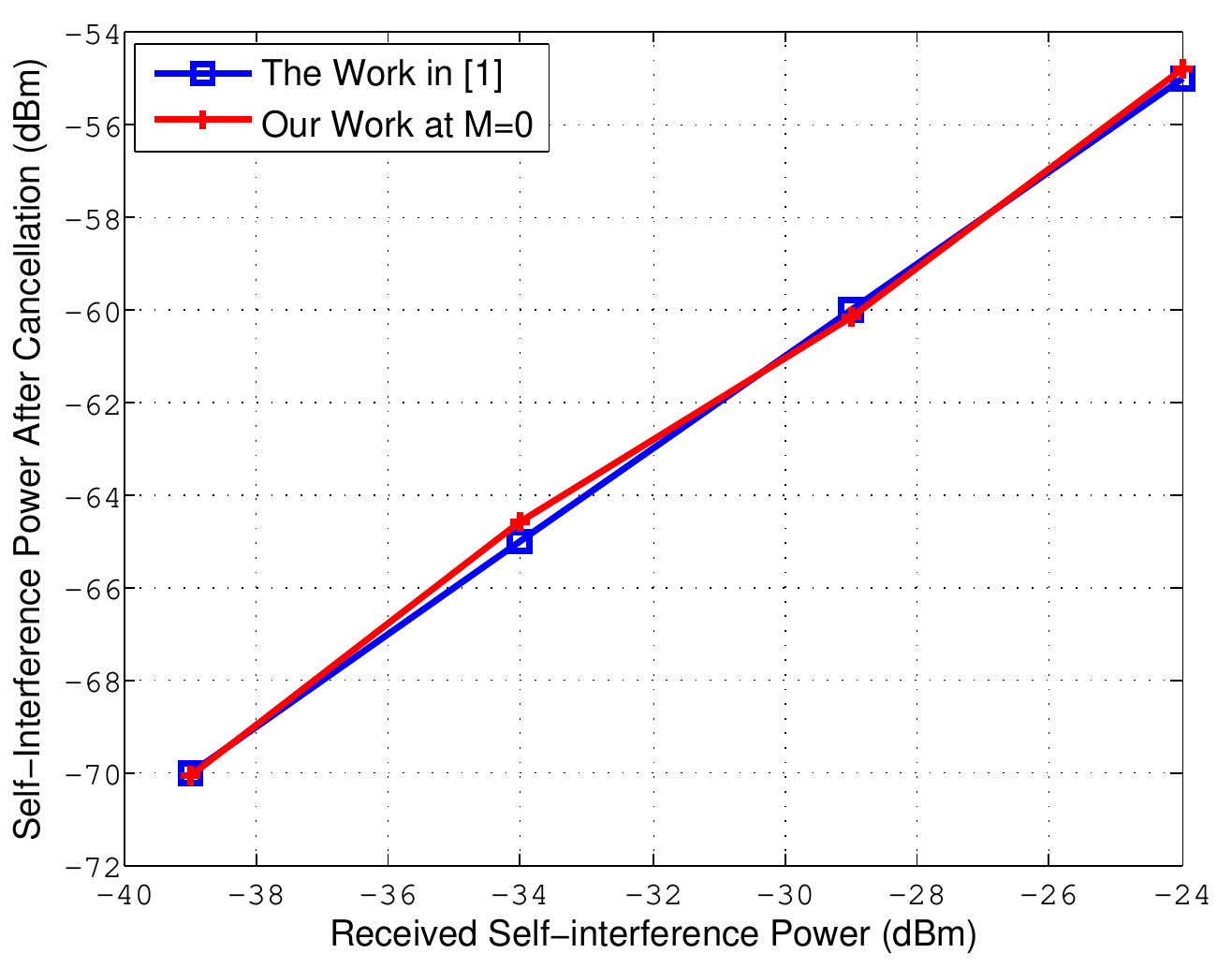}
  \caption{Experimental results at $M=0$ compared to the experimental results in~\cite{Ref1}.\label{Fig10Label}}
\end{center}
\end{figure}

First, in order to validate our experimental setup, our results are compared to the experimental results reported in~\cite{Ref1}. In~\cite{Ref1} the WARP platform is used to characterize the cancellation capability of different self-interference cancellation techniques under different antenna configurations. The cancellation techniques used in~\cite{Ref1} only consider the suppression of the CPE component, therefore, it has to be compared with our results at $M=0$. Figure~\ref{Fig10Label} shows the cancellation performance of the digital self-interference cancellation technique in~\cite{Ref1} compared to our digital self-interference cancellation technique at $M=0$.
\begin{figure}[!ht]
\begin{center}
\noindent
  \includegraphics[width=6.5in ,height=3in]{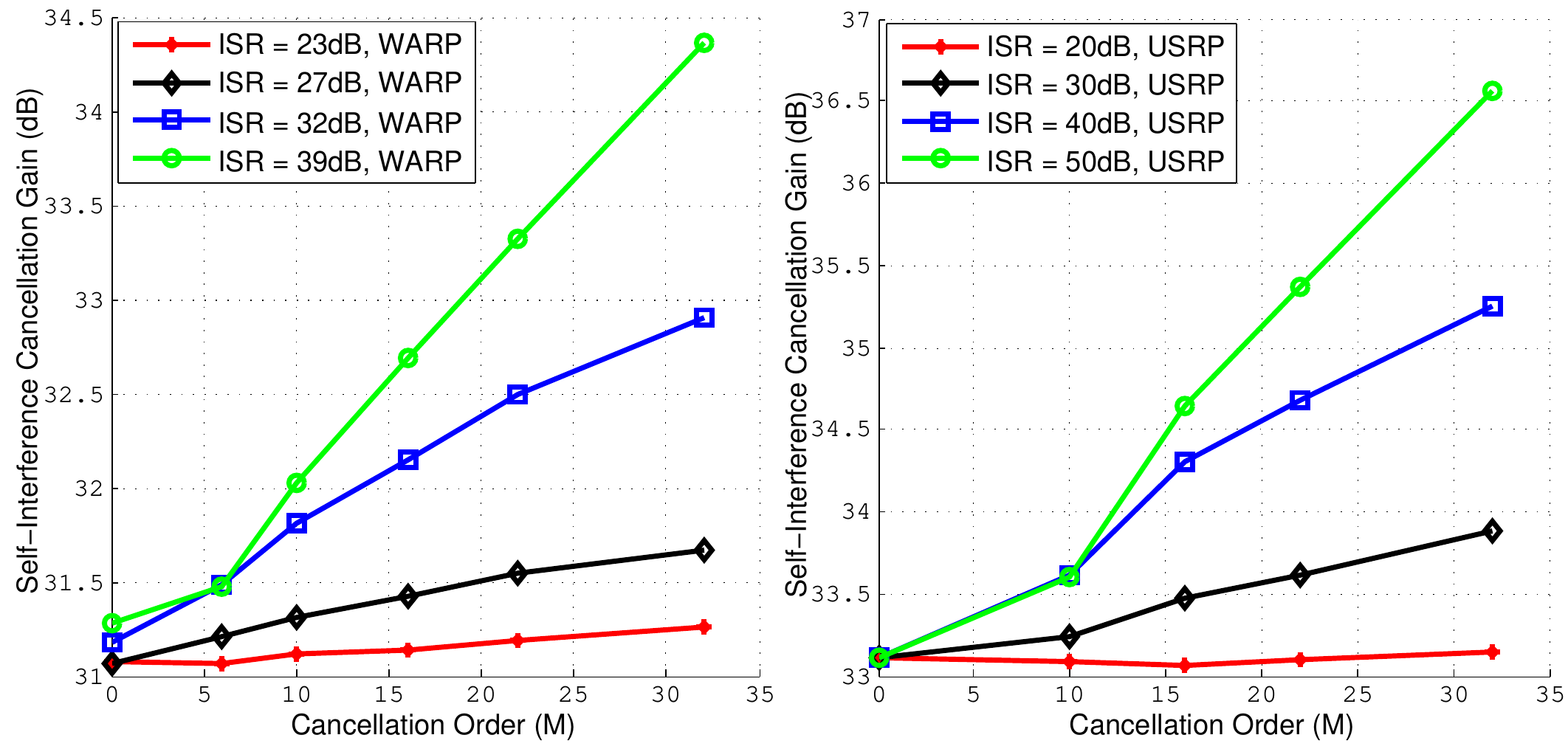}
  \caption{Self-interference cancellation gain for frequency-domain phase noise suppression technique using WARP and USRP platforms.\label{Fig11Label}}
\end{center}
\end{figure}
%
%
%
%
\begin{figure}[!ht]
\begin{center}
\noindent
  \includegraphics[width=4in ,height=3in]{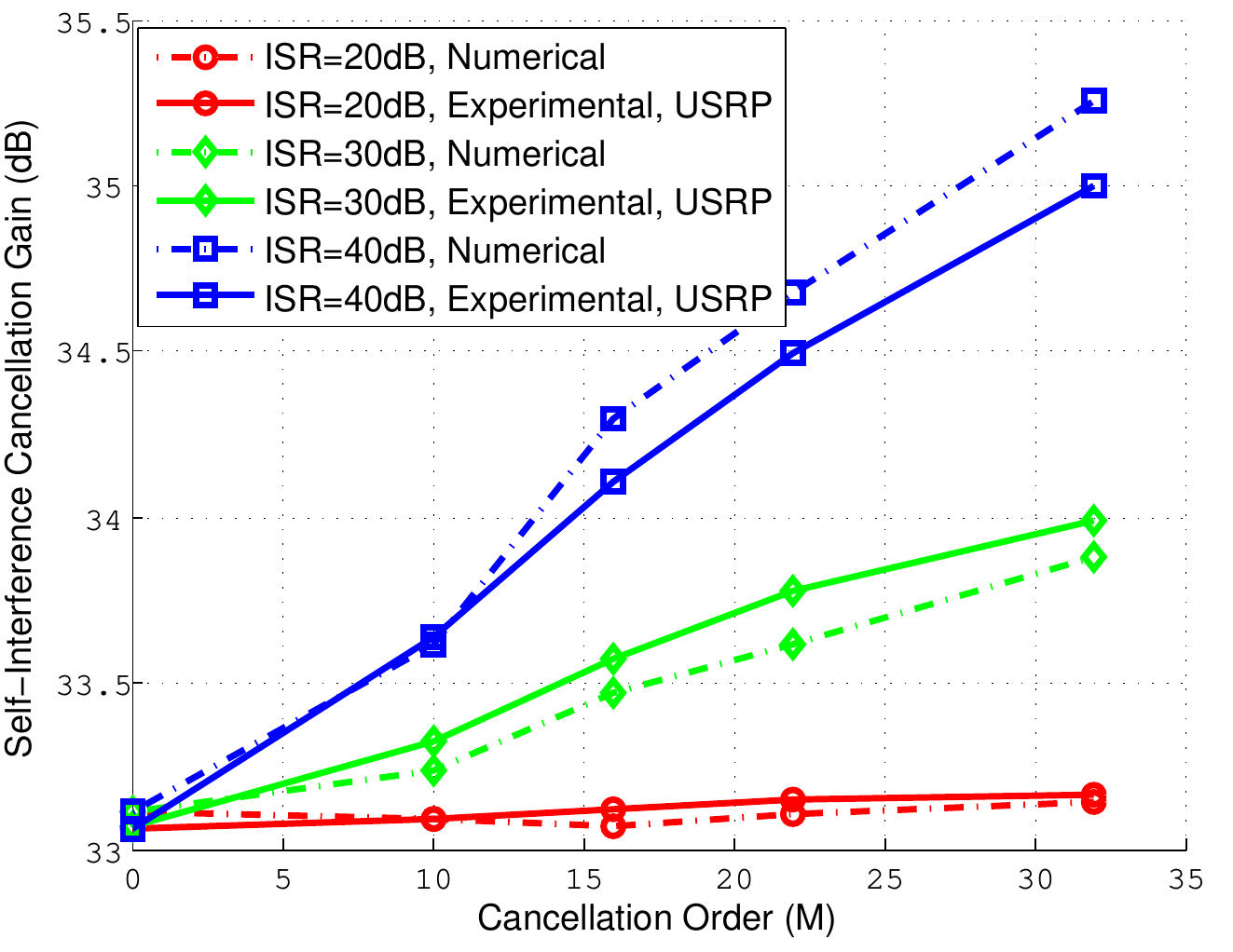}
  \caption{Experimental Vs numerical results for frequency-domain phase noise suppression at $P_{ICI} = -33$dBc, using USRP platform.\label{Fig13Label}}
\end{center}
\end{figure}

Now we investigate the performance of the proposed phase noise estimation and suppression techniques. Figure~\ref{Fig11Label} show the achieved self-interference cancellation gain at different ISR values for the proposed frequency-domain cancelation technique running over WARP and USRP platforms respectively. The results confirm the conclusions derived from the numerical results, where performing ICI suppression improves the cancellation performance (up to 3dB more cancellation) especially at high ISR values. The results also show that the USRP platform achieves better performance than the WARP platform, which means that the quality of the RF circuits and connectors used in the USRP may be better than those used in the WARP platform.

In order to measure the accuracy of the numerical analysis presented earlier in this section, the system model is simulated using the same system parameters used in the USRP platform, and the results are then compared to the experimental results. As shown in Figure~\ref{Fig13Label}, the simulation results highly matches the experimental results with $<$0.2dB error. With this matching, all the conclusions derived from the numerical analysis are now confirmed.
\subsection{Overall system performance and discussions}
Generally, as shown in the previous analysis, phase noise suppression could achieve a maximum of 6dB more self-interference cancellation with free-running oscillators (reduced to 3dB in case of PLL based oscillators), which is considered a small gain compared to the required computational complexity ($O(M^2), M=32$). Not only that, but this gain is further reduced with the decrease of the ISR. Three main reasons explain the low gain achieved, and the high complexity required for phase noise estimation and suppression in full-duplex systems; first, the most important reason is that phase noise in full-duplex systems has to be estimated in the presence of the unknown signal-of-interest, which significantly affects the estimator performance, especially at high signal-of-interest powers. Second, phase noise is a fast time-varying process that needs to be estimated at every sample and at every subcarrier (i.e. high complexity). Finally, in practical systems where PLL based oscillators are used, and due to the phase noise spectrum flatness, achieving high suppression gains require the phase noise coefficients to be estimated at all OFDM symbol subcarriers, which results in significant complexity especially in OFDM systems with a large number of subcarriers.

The previous discussion raises the following important questions; In practical scenarios does phase noise suppression improve the overall system performance? If yes, by how much?, and is it worth the extra complexity? In order to answer these questions, we investigate the overall system performance of a common practical system in typical operating conditions.

\begin{figure}[!ht]
\begin{center}
\noindent
  \includegraphics[width=4in ,height=3in]{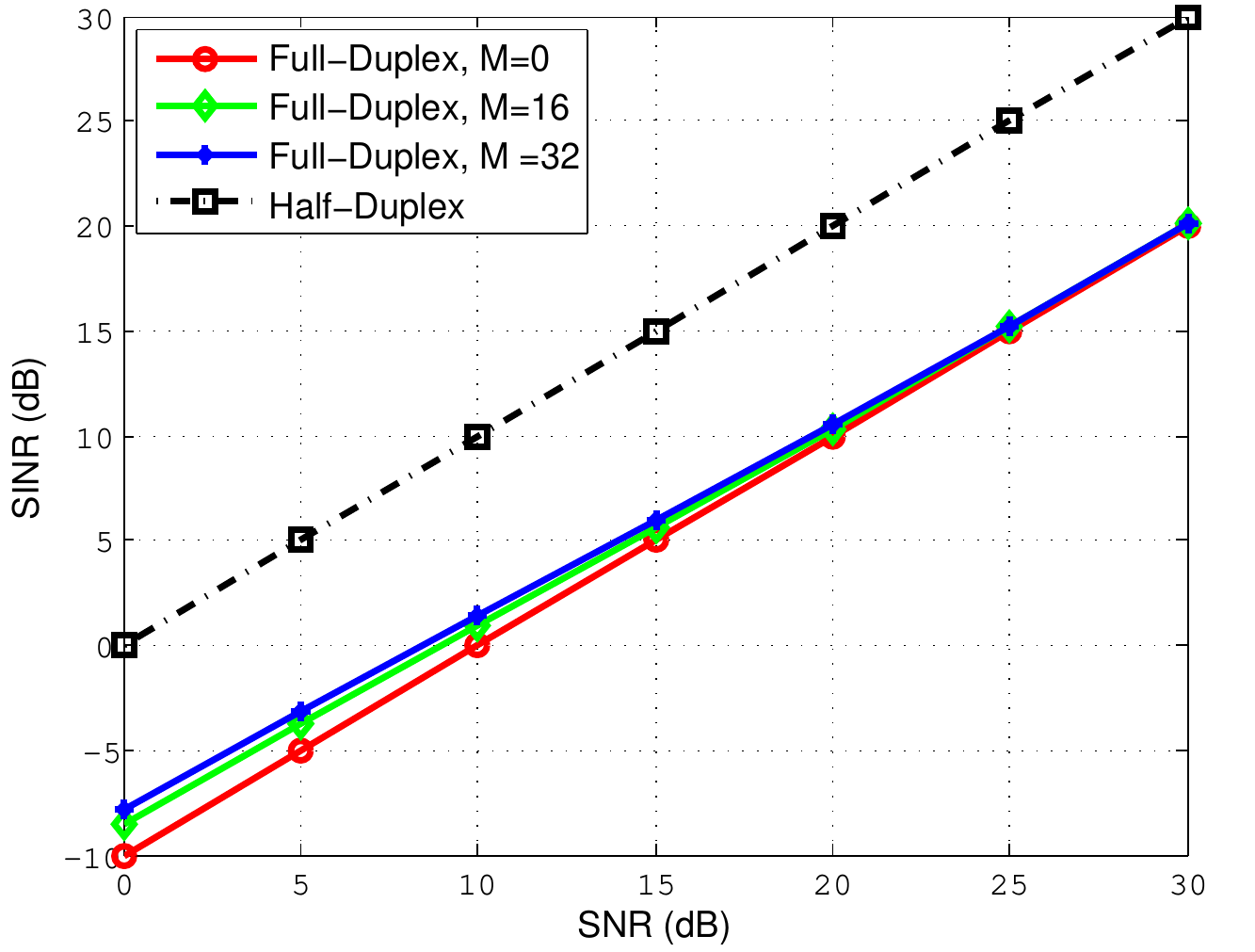}
  \caption{Overall SINR for time-domain phase noise suppression technique at different SNR values with 60dB passive self-interference suppression and $P_{ICI}$ of $-$40dBc, using PLL based oscillator and estimated channel.\label{Fig14Label}}
\end{center}
\end{figure}

Assume a 20MHz WiFi system with a 20dBm transmit power, $-$90dBm noise floor, and $-$40dBc total in-band phase noise power (i.e. $P_{ICI}=-40$dBc). As discussed in section II, in practical full-duplex systems, a combination of different self-interference cancellation techniques is used. In our example, a 60dB combined passive and RF self-interference suppression gain is assumed~\cite{Ref7}. Following the passive and RF suppression, the proposed self-interference cancellation and phase noise suppression techniques are used for digital cancellation. After self-interference cancellation, the overall signal to interference plus noise ratio (SINR) is calculated. The SINR is then used as a performance metric to characterize the overall system performance. The full-duplex system performance is compared to the corresponding half-duplex system performance at different signal to noise ratios (SNR). The SNR is defined as the signal-of-interest power divided by the noise floor (i.e. SNR of the corresponding half-duplex system). Figure~\ref{Fig14Label}, shows the overall SINR for both half-duplex and full-duplex systems. The performance is evaluated using a PLL based oscillator. The channel information is estimated at the beginning of each transmission frame.

In this example, the signal is transmitted at 20dBm power; the RF cancellation has 60dB gain, which results in a received self-interference power of -40dBm (after passive suppression). The ICI power is $-$40dB below the self-interference, that is $-$80dBm. The system is simulated at SNR ranging from 0 to 30dB, which corresponds to signal-of-interest power ranging from $-$90dBm to $-$60dBm respectively (the noise floor is $-$90dBm). First, without ICI suppression ($M=0$), the remaining self-interference power will be limited by the ICI power level ($-$80dBm in this case). As a result, the full-duplex system noise floor will be increased to $-$80dBm, resulting in a 10dB difference in SINR between full- and half-duplex performance. The results in Figure~\ref{Fig14Label} also show that the maximum ICI suppression gain is achieved at low SNR scenarios; this is where the ICI power is greater than the signal-of-interest power. However, the achieved suppression gain (e.g. $\sim$2.5dB) is not enough to fill the gap between the full-duplex and half-duplex performance.

As a conclusion, we see that phase noise is a main performance limiting factor in full-duplex systems. Generally, phase noise power could be reduced either by designing good quality oscillators, or using estimation and suppression techniques such as the proposed techniques. Phase noise has two main effects; CPE and ICI. Elimination of the CPE component is essential and very simple. However, the elimination of the ICI component is a very challenging and complexity consuming process. The gain achieved by performing ICI suppression is relatively small compared to the required complexity. Furthermore, the ICI suppression gain decreases significantly at high SNR scenarios, mainly due to the presence of the signal-of-interest during the estimation process. On the other hand, using free-running oscillators approximately double the achieved ICI suppression gain compared to PLL based oscillators. However, due to its continuous phase drift, free-running oscillators might be useful only in short packet systems.

\section{Conclusions}
In this paper, the problem of phase noise estimation and suppression in OFDM full-duplex systems is analytically and experimentally investigated. The effect of phase noise on full-duplex systems is studied showing that phase noise in full-duplex systems causes two main effects; CPE and ICI. A frequency-domain and a lower complexity time-domain ICI suppression techniques are proposed. The feasibility of performing ICI suppression in full-duplex systems is investigated in terms of required complexity and achieved gain. Both free-running and PLL based oscillators are considered. The results show that ICI suppression in OFDM based full-duplex systems is a very challenging and complexity consuming problem. More specifically, the results show that at a complexity of order $O(32^2)$ a maximum of 6dB more self-interference cancellation is achieved compared to the case where no ICI suppression in performed. This gain is reduced to 3dB when PLL based oscillator is used, mainly due to the flatter spectrum of the phase noise in PLL oscillators. Furthermore, the results show that this gain is conditioned; it can be achieved only at low SNR scenarios where the phase noise power is greater than the signal-of-interest power. However, at high SNR scenarios, ICI suppression does not add any gain. As a conclusion, since phase noise is one of the main limiting factors in full-duplex systems, phase noise suppression should be considered. Phase noise suppression could be achieved by either designing high quality oscillators or using estimation and suppression techniques. However, the analysis in this paper show that the gain achieved by phase noise suppression is relatively small compared to required complexity, especially in practical systems where PLL based oscillators are used.
\section{Acknowledgment}
The authors would like to express their sincere appreciation to Professor Ashutosh Sabharwal with the Department of Electrical and Computer Engineering, Rice University, Houston, TX, for the numerous insightful discussions regarding full duplex systems, their applications and limitations .
%


\end{document}